\documentclass{article}

\usepackage{PRIMEarxiv}

\usepackage[utf8]{inputenc} 
\usepackage[T1]{fontenc}    
\usepackage{hyperref}       
\usepackage{url}            
\usepackage{booktabs}       
\usepackage{amsfonts}       
\usepackage{nicefrac}       
\usepackage{microtype}      
\usepackage{lipsum}
\usepackage{fancyhdr}       
\usepackage{graphicx} 
\usepackage{multirow}
\usepackage{float}
\graphicspath{{media/}}     
\usepackage{lineno}

\title{Impacts of Census Differential Privacy for Small-Area Disease Mapping to Monitor Health Inequities}

\author{%
  Yanran Li$^1$, Brent A. Coull$^1$, Nancy Krieger$^2$, Emily Peterson$^3$,\\\textbf{Lance A. Waller$^3$, Jarvis T. Chen$^2$, Rachel C. Nethery$^{1*}$} \\
  $^1$Department of Biostatistics, Harvard TH Chan School of Public Health, Boston, MA \\
  $^2$Department of Social and Behavioral Sciences, Harvard TH Chan School of Public Health, Boston, MA\\
  $^3$Department of Biostatistics and Bioinformatics, Emory Rollins School of Public Health, Atlanta, GA\\
  $^*$Corresponding author
}

\begin{document}
\maketitle
\section*{Abstract}
The US Census Bureau will implement a new privacy-preserving disclosure avoidance system (DAS), which includes application of differential privacy, on the public-release 2020 census data. There are concerns that the DAS may bias small-area and demographically-stratified population counts, which play a critical role in public health research and policy, serving as denominators in estimation of disease/mortality rates. Employing three DAS demonstration products, we quantify errors attributable to reliance on DAS-protected denominators in standard small-area disease mapping models for characterizing health inequities. We conduct simulation studies and real data analyses of inequities in premature mortality at the census tract level in Massachusetts. Results show that overall patterns of inequity by racialized group and economic deprivation level are not compromised by the DAS. While early versions of DAS induce errors in mortality rate estimation that are larger for Black than for non-Hispanic white populations, this issue is ameliorated in newer DAS versions.

\subsection*{Teaser}
In simulated and real data, we evaluate DAS-affected population counts as denominators in small-area disease rate estimation.

\section{Introduction}
 \label{intro}

In order to release statistics about populations without violating privacy, the US Census Bureau announced in 2018 that it would implement differential privacy (DP) on publicly released data products derived from 2020 decennial census data\cite{doi:10.1073/pnas.200371411}. Given that reporting of statistics at population-scale is not sufficient to ensure the privacy of individuals\cite{10.1007/11787006_1}, DP provides a formal framework for providing rigorous privacy guarantees via the principled introduction of noise into statistics. That is, DP ensures that at specified population-scales the statistics do not change substantially when a single individual’s information is included/excluded. The Census Bureau's use of DP as a component of its 2020 Disclosure Avoidance System (DAS), which was introduced in response to a report showing that its previous methods permitted larger than expected risks of person re-identification\cite{nationalacademies.org}, is a departure from the disclosure avoidance procedures applied to prior publicly released decennial census data (described by \cite{hotz2022chronicle}). The variant of DP applied to the 2020 census products is a top-down mechanism that infuses random noise into census tabulations at six different nested geolevels (nation, state, county, tract, block group, block)\cite{garfinkel2019deploying,abowd2022tda}. Because the threat to privacy is greater for release of statistics on small populations, the noise injected into small counts is relatively larger than that applied to larger counts\cite{cohen2022private, https://doi.org/10.1111/1475-6773.14000}.

The decision to incorporate DP and the necessary subsequent post-processing steps in the 2020 Census DAS has been controversial. Some scholars have voiced concerns about the potential negative impacts of noisy data on public policy and social science research, which critically rely upon census data\cite{doi:10.1126/sciadv.abk3283}.  What's more, there are concerns about impacts on resource allocation, since many programs rely on population counts. Harvard Data Science Review (HDSR) recently released a special issue to document, contextualize, and assess US Census Bureau’s adoption of DP and presented discussions with key stakeholders about the decision\cite{Gong2022Harnessing}. Within this issue is a guide for researchers on responsible use of 2020 public-release decennial census products\cite{groshen2022disclosure}.

Monitoring of social and spatial patterns of disease, a fundamental component of public health research and practice, also requires accurate population counts, which serve as the denominator for estimation of disease rates, and these population denominators are most often obtained from census products. In particular, disease monitoring typically relies on demographically-stratified small-area disease rates (and associated population denominators), which are analyzed to identify and intervene on populations at highest risk. Moreover, populations within small areas tend to be more homogeneous than in larger areas (reflecting impacts of present and past residential racialized and economic segregation and housing costs)\cite{rothstein2017color,trounstine2018segregation,widestrom2015displacing,ellen2019dream,orlando2021keeping}, providing a differential between the socioeconomic and environmental characteristics of areas studied that may aid in detecting relationships between these variables and health data\cite{piel2020small}. Thus, the potential exaggerated effects of DP on census tabulations for small populations is of particular concern to the public health community.

Although the US Census Bureau will solely release the DAS-protected 2020 Census tabulations, they have published several "demonstration products" in which sequentially refined variants of the proposed DAS procedures were applied to publicly released 2010 decennial census data to collect public comment\cite{2019dp,0527dp}. Through comparisons with the original 2010 public release census data, these products enable researchers and practitioners to assess whether DP adjustments unintentionally induce systematic (instead of random) discrepancies in reported Census statistics or in analyses that utilize them. Scholars across a range of fields have thus evaluated the DAS-affected products and given feedback to the Census Bureau, which has led to modifications and shaped the trade-offs between "accuracy" and "privacy" in the final 2020 DAS\cite{doi:10.1126/sciadv.abk3283}. In total, three different demonstration products have been released (with increasingly refined DAS procedures applied in each new product) that provide the race and age stratified population counts needed for small-area studies of health inequities.

While the demonstration products have been used extensively to assess the potential impacts of the 2020 DAS on redistricting for political representation\cite{kenny2021use,cohen2021,cohen2022private}, few studies have formally assessed its impacts in public health applications. Even fewer have focused on the consequences of using DP population counts in modeling of small-area disease rates for identifying health trends and monitoring disparities. One recent study compared county-level estimates of 2010 racialized group mortality rates produced using the original 2010 census population counts and the first of the Census Bureau's demonstration products\cite{doi:10.1073/pnas.200371411}. This work estimated county-level rate differences for rates constructed with the original and demonstration product denominators and summarized the rate differences across county urbanicity strata. Using a similar approach but reporting county-level absolute percent errors in mortality rate estimates (comparing demonstration product vs. original denominators), another study investigated the extent to which DP could distort county-level COVID-19 mortality rates by age-sex/racialized groups\cite{doi:10.1177/2378023121994014}. They employed the second of the demonstration products and concluded that DAS-induced errors in COVID-19 mortality rates were larger for non-white racialized groups (though their use of absolute errors precludes assessment of directionality of errors). Another study compared estimates of premature mortality rates aggregated by racialized group and by census tract quintile of inequality across Massachusetts using denominators from the 2010 census vs. the first DAS demonstration product. Although census tract inequality measures were used to create strata (each stratum included $>200$ census tracts), all mortality rates were computed and compared in aggregate for the strata, i.e., the study did not evaluate DAS impacts on small area rate estimates. They concluded that, for these heavily aggregated metrics, the 2020 DAS procedures may have little impact on estimates of health inequities\cite{Krieger2021}. Given the limited nature of this literature, numerous gaps remain. For instance, none of these studies have (1) examined and compared different DP versions' impacts for estimation of disease rates for small areas (smaller than counties) commonly used for health disparities analysis in public health practice; (2) utilized the March 2022 released demonstration product that better reflects the final 2020 DAS procedures; nor (3) conducted simulation studies to more formally quantify biases introduced by DP.

Therefore, in this paper we leverage all three of the demonstration products available at this time to evaluate the performance of small-area disease mapping models employing DAS-affected denominators vs. original 2010 decennial census denominators, with an emphasis on accurate characterization of health inequities. Potential biases are illustrated using a pseudo-simulation study and a real data analysis of racialized disparities in premature mortality at the census tract (CT) level in Massachusetts (MA). Our results may help public health researchers and practitioners to determine whether the 2020 DAS-affected publicly released products will yield reliable and actionable results when used for essential public health tasks. 

\section{Materials and Methods}
\label{methods}

\subsection{Population count data}

To help data users study impacts of the proposed 2020 DAS, the Census Bureau has released a set of demonstration data products in which potential variants of the 2020 DAS have been applied to 2010 decennial census (DC) data. Data available include 2010 population counts both without and with DP applied (along with subsequent post-processing), enabling comparisons of the population counts themselves and the results of analyses that rely on them. Here, we use three of these products to evaluate how using new DAS-protected population counts to model disease rates might bias our results if the true population counts are those from the 2010 DC (see Section 4 for discussion of how DC population counts themselves are known to exhibit systematic biases). In both the simulations and real data analysis, we use the following four CT level population count data sources for MA: 
\begin{enumerate}
    \item \textit{Original 2010 DC Data}\cite{dc};
    \item \textit{Demonstration product released in 2019 (DP19): 2010 DC data with the first version of Census Bureau's DAS procedure applied}\cite{2019dp};
    \item \textit{Demonstration product released in 2020 (DP20): 2010 DC data with the second version of Census Bureau's DAS procedure applied}\cite{0527dp};
    \item \textit{Demonstration product released in 2022 (DP22): 2010 DC data with the third version of Census Bureau's DAS procedure applied}\cite{2022dp}.
    
\end{enumerate}

For the original DC data, we use the \texttt{tidycensus} R package\cite{walker2020tidycensus} to extract CT population counts stratified by age and census-defined “racial” and “ethnic” categories (which we refer to as racialized groups\cite{krieger2000counting}, since these categories are socially constructed) for the state of MA. In our study, we consider these to be “ground truth” population counts. For the three demonstration data products, we obtain population counts stratified by age and racialized group for MA CTs from the website of IPUMS (Integrated Public Use Microdata Series) National Historical Geographic Information System (NHGIS)\cite{IPUMS}.

\subsubsection{Differences in demonstration products}
In the differential privacy algorithm, the accuracy privacy-loss tradeoff is controlled by the privacy-loss budget (PLB) parameter, $\epsilon$, representing the spectrum between perfect privacy/low accuracy ($\epsilon = 0$), to perfect accuracy/low privacy ($\epsilon = \infty$). 
The demonstration products of DP19 and DP20 considered here use the same value of $\epsilon$ ($\epsilon=6.0$ overall, divided between the population tables, $\epsilon=4.0$, and housing and household tables, $\epsilon=2.0$), and thereby identical implementations of differential privacy. 

The difference between DP19 and DP20 lies only in the post-processing procedures, which are operations involving how the DAS TopDown Algorithm (TDA) converts the formally private noisy tabulations taken from the confidential data into the non-negative integer counts that will be published. The TDA used in DP19 conducted the postprocessing of all of the statistics for a particular geographic level at the same time, resulting in distortions when there were large quantities of statistics with zeros or very small values processed at the same time. To address and mitigate this issue, the TDA used in DP20\cite{PPMF0527factsheet} conducts the postprocessing in a series of passes through all the geographic levels (national level, state level, etc.). Specifically, the first pass processed total population counts, and the second pass processed statistics necessary to inform redistricting. The third pass processed core statistics stratified by age/sex/racialized group, and the final pass processed all remaining counts. In this version of the TDA, output from each pass was constrained to agree with the counts from prior passes.

DP22, on the other hand, incorporates modifications to the DP algorithm parameters in response to stakeholder feedback that greater accuracy was needed. Specifically, the Census Bureau tuned the PLB applied to different sets of tabulations\cite{progress22}. The PLBs assigned to person-level and housing unit-level counts in DP22 are $\epsilon =20.82$ and $\epsilon =22.77$, respectively\cite{factsheet2022}, both of which are substantially higher than the analogous PLBs for DP19 and DP20, yielding counts with lower privacy/higher accuracy. DP22 applies the same multi-pass post-processing procedures as DP20, but incorporates additional geographic entities into this post-processing\cite{factsheet2022}. Importantly, DP22 may not represent the final version of DAS procedures that will be applied to the 2020 census data, as decisions about the final version have not been announced at the time of writing. However, DP22 is the recently released demonstration product and reflects the recent refinements of the algorithm at this time.

\subsection{Real data analysis of inequities in premature mortality}

The outcome of interest of our study is premature mortality (death before 65 years old), which is an important and widely-used metric in health inequities studies. We focus on inequities in risk by racialized group (comparing Black and non-Hispanic White (NHW) populations) and by socioeconomic status. We obtained records of all premature deaths in 2010 from the MA Department of Public Health\cite{MADPH}. These data have been described previously, see Krieger et al.\cite{Krieger2021} for more detail. Briefly, each record contains the age, racialized group, and residential address for the deceased individual. We geocoded the addresses to CTs and created aggregate premature mortality counts stratified by age group (for ages <65), race/ethnicity, and CT.

Using each of the four population count data sources described above, we compute age-standardized 2010 premature standardized mortality ratios (SMR) for both the Black and NHW populations in each MA CT using the indirect standardization method (described below)\cite{boscoe2013geographic}. The SMRs are calculated using the CT/racialized group observed count in the numerator and an expected premature mortality count for the CT, based on its population size and age distribution, in the denominator.

\subsubsection{Age-standardization}

To perform indirect age-standardization and create expected counts to be used as denominators based on the DC, DP19, DP20, and DP22 population counts, we use the \textit{ageadjust.indirect()} function from the \texttt{epitools} R package\cite{epitools}. Age-standardization adjusts for differences in age distribution to mitigate possible confounding effects on inequity analyses arising from differing age distributions across groups. We conduct age-standardization, based on empirical MA statewide age group-specific premature mortality rates, separately for each CT and racialized group, to get expected premature mortality counts. We compute descriptive statistics for the expected counts constructed from each dataset (DC, DP19, DP20 and DP22), and we specifically compare the DC expected counts, used here as the ``ground truth'', with the DP19, DP20 and DP22 expected counts, which have the DAS applied. 

\subsubsection{Models}

Our four sets of CT-level SMRs correspond to the four different sets of denominator data (DC, DP19, DP20, and DP22). To study the impact of the different denominators in assessing disparities in premature mortality, we fit standard disease mapping models to the CT-level SMRs stratified by racialized group from each denominator source (separately) to examine associations with racialized group and an area-level measure of economic deprivation -- CT proportion below the poverty line (PropPov)-- extracted from the 2008-2012 5-year American Community Survey (ACS) data, which is estimated for the CT as a whole and is not race-specific.

The four sets of SMRs are modeled separately, using a multi-level variant of the spatial Poisson regression model\cite{BesagJ1991}: Let $i = {1,\cdots,N}$ index CTs in MA and $j = {0, 1}$ index racialized group (0 for NHW and 1 for Black), so that $Y_{ij}$ is the premature mortality count in racialized group $j$ within CT $i$. $I(Black)_{ij}$ is a binary indicator of Black racialized group, and $PropPov_i$ is the proportion in poverty in CT $i$ (centered and scaled). Let $P_{ij}$ be the CT- and racialized group-specific expected number of premature mortalities computed using any of the four datasets described above. We assume $Y_{ij} \sim Poisson(\lambda_{ij})$ and fit the following model using each of the four variants of $P_{ij}$ formed from DC, DP19, DP20, and DP22:
\begin{equation}
    log(\lambda_{ij}) = \beta_0 + \beta_1I(Black)_{ij} + \beta_2PropPov_{i} + \theta_i + \phi_{ij} + log(P_{ij}),
\end{equation}

where $\theta_i$ is a CT-specific random effect with a conditionally autoregressive spatial covariance structure\cite{10.1007/978-1-4612-1284-3_4} and $\phi_{ij}$ is an unstructured CT- and racialized group-specific error term that models overdispersion in the disease count. To clarify, $\theta_i$ is spatial random intercept unique to CTs but shared by racialized groups within a CT, while the $\phi_{ij}$ are both CT and racialized group-specific. Models are fit by a Bayesian approach implemented in the \texttt{CARBayes} package in R\cite{JSSv055i13}. For each of the four denominator data sources, mortality rate ratio (MRR) estimates based on the exponentiated posterior means of the coefficients and 95\% credible intervals are reported and compared in Section \ref{real pmr}.

\subsection{Simulation Study}

In addition to the real data analysis, we conduct a simulation study to formally assess the magnitude of biases induced in estimates of health inequities due to using the DAS-protected denominators in standard models. We structure our simulated outcomes to mimic real patterns in premature mortality in MA. Synthetic premature mortality counts are generated for each CT, stratified by racialized group (NHW and Black), using the 2010 DC expected premature mortality counts as the denominator and the real covariate data, and following the model form described in the real data analyses above. We then fit models to the simulated data using the DP19, DP20 and DP22 expected counts, but otherwise correctly specified, and evaluate the resulting bias in key parameters. Further details are given below.

\subsubsection{Data Generating Process}
Using the 2010 DC expected counts for each CT and racialized group, we simulate the outcomes. Formally, premature mortality counts, $Y$, are generated following the model form in equation (1), plugging in the real DC-based expected counts for $P_{ij}$ and using the real CT-level PropPov variable. The coefficient parameter values used in all simulations are $\beta_0=0$, $\beta_1=0.4$, and $\beta_2=0.01$.

The conditionally autoregressive spatial effect $\theta_i$ is generated as $$\theta_i |\theta_{-i} \sim N\left(\frac{0.2\sum_{k}w_{ik}y_k}{w_{i+}}, \frac{1}{w_{i+}}\right),$$ where $w_{ik}$ is the $(i, k)^{th}$ element of an adjacency matrix $W$, and $w_{i+}$ is the sum of the elements in the $i^{th}$ row of $W$. $\phi_{ij} \sim N(0, 0.25)$ is an unstructured random effect. The hyperparameter values in the distributions of $\theta_i$ and $\phi_{ij}$ were selected to generate data with moderate spatial correlation and an outcome distribution mimicking the empirical distribution of CT premature mortality rates in MA.

We simulate 100 datasets from this model, and for each simulated dataset we fit four models to it-- one plugging in each set of expected counts (DC, DP19, DP20 and DP22) as the denominator. Aside from possible error in the denominators, the fitted models are otherwise correctly specified, to allow us to isolate potential biases due to DAS-induced error in the denominators.

\subsubsection{Model Assessment}
To investigate the performance of the models fit with different denominator data sources, we evaluate the distribution of the estimated model coefficients and the model-based SMR estimates, relative to the known true values of these quantities. For each coefficient, we compute the simulated bias of the estimator based on each of the four models using different denominator sources. We summarize and visualize these coefficient-specific estimates across all 100 simulated datasets to demonstrate how the use of DAS-protected denominators in model fitting (when DC denominators are the "true" denominators in the data generating process) impacts assessment of high-level patterns and comparisons of risks across groups.

For a given simulated dataset (indexed by $k=1,...,100$)  and denominator data source $(P_{ij})$, we estimate the model-based SMRs as:
\begin{equation}
    \widehat{SMR}_{ijk} = \frac{\widehat{Y}_{ijk}}{P_{ij}},
    \label{smr_dp}
\end{equation}
where $\widehat{Y}_{ijk}$ is the predicted value of $Y_{ij}$ from the model fit to simulated dataset $k$ using the given denominator data source. We then compute the bias and mean absolute percentage error (MAPE) for each CT and racialized group's SMR estimate for each denominator data source, i.e.,

\begin{equation}
    MAPE_{ij} = \frac{1}{100}\sum_{k=1}^{100}\left|\frac{\widehat{SMR}_{ijk}-(\lambda_{ijk}/P_{ij})}{(\lambda_{ijk}/P_{ij})}\right|,
    \label{mape}
\end{equation}
\begin{equation}
    Bias_{ij} = \frac{1}{100}\sum_{k=1}^{100}\left(\widehat{SMR}_{ijk}-(\lambda_{ijk}/P_{ij})\right).
    \label{bias}
\end{equation}

We plot and map these metrics for the DC, DP19, DP20, and DP22 denominators (for the racialized groups separately). This enables us to investigate whether small-area spatial patterns in model-smoothed disease/mortality risk estimates are preserved when using the DAS-protected denominators in standard models. In this way, we can assess extent and direction of biases related to DP algorithm.

\section{Results}
\label{results}

\subsection{Comparison of denominator data sources}
Figure \ref{scatterplot} in the Appendix shows a scatterplot of the racialized group-stratified CT expected premature mortality counts from the DC vs. DP19, DP20 and DP22. From this figure, it is clear that that Black expected counts are generally much smaller than the NHW counts (Black individuals represented about 7.5\% of the MA population in 2010) and that for both racialized groups, the DP19 data are slightly more noisy than the DP20 data, which are more noisy than the DP22 data. The mean (and standard deviation) of the differences in the CT-level DP and DC expected counts for the NHW population are $0.0029$ (0.562) for DP19, $0.0012$ (0.484) for DP20 and $0.0007$ (0.033) for DP22 and for the Black population are $0.0007$ (0.139) for DP19, $-0.0003$ (0.105) for DP20 and $0.0001$ (0.022) for DP22. This again demonstrates that the DP22 data have less bias and less noise, on average, than the DP20 data, which also have less bias and less noise than the DP19 data. The smaller magnitude of bias and noise associated with Black expected counts relative to NHW in both DP datasets is a result of the smaller scale of the Black counts. In Figure \ref{boxplot_percent_error}, we plot the percent error in the DP19, DP20 and DP22 CT expected premature mortality counts, relative to the DC expected counts (the ``ground truth''), stratified by race group. First, we note that the distribution of percent errors in the DAS-protected expected counts for the NHW population is narrow and centered around zero, a result of the generally large NHW populations in most MA CTs. Second, the distribution of percent errors in DAS-protected expected counts for Black populations is much wider than for NHW, a result of the generally small Black populations in most MA CTs. Moreover, for DP19 and DP20, the distribution of percent errors for Black populations is centered well below zero, indicating that Black expected counts tend to be underestimated in the earlier DAS variants. To more thoroughly characterize this under-estimation as well as the comparison between NHW and Black, we include in Table \ref{expected_counts_percent-table} the percent of expected counts that are under-estimated for each demonstration product (relative to the DC). In DP22, with the increased PLB, the distribution of errors for Black populations remains wider than for NHW but is centered around zero.

\subsection{Real Premature Mortality Modeling Results}
\label{real pmr}
MRR estimates and 95\% Bayesian credible intervals from our racialized group-stratified models are presented in Table \ref{IRR-table}. The racialized group variable is a group-level binary indicator of Black (versus NHW) and the MRR estimate for the racialized group variable is $>1$ for the models fit with each of the four denominator data sources, indicating that on average CT premature mortality rates for Black populations are higher than rates for NHW populations. Percent of CT residents in poverty is also associated with higher premature mortality rates in the racialized group-stratified models. Inferences about patterns of health disparities from the models using the four different denominator sources are identical, with only very minor differences in the point estimates (and even these may be attributable to randomness in the Bayesian posterior sampling).

\subsection{Simulation Results}

\begin{figure}[H]
\caption{Boxplots of the percent error in DP19, DP20 and DP22 CT expected premature mortality counts, relative to DC expected counts (the ``ground truth''), for Black and non-Hispanic white populations.} 
\centering 
\includegraphics[width=0.8\textwidth]{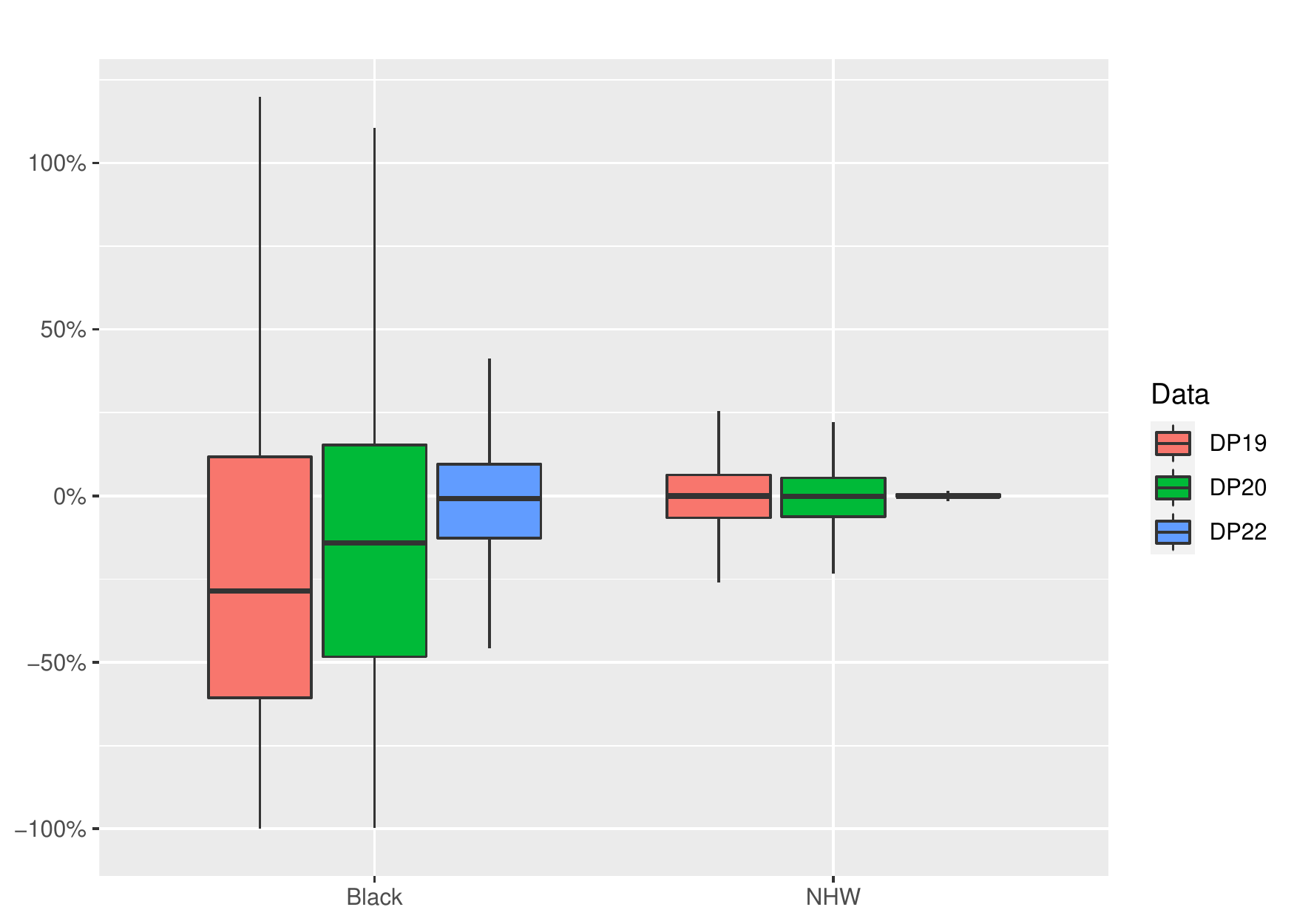} 
\label{boxplot_percent_error}
\end{figure}

\begin{figure}[H]
\caption{Boxplot of estimated coefficients' biases across simulations using the four different denominator data sources.} 
\centering 
\includegraphics[width=0.9\textwidth]{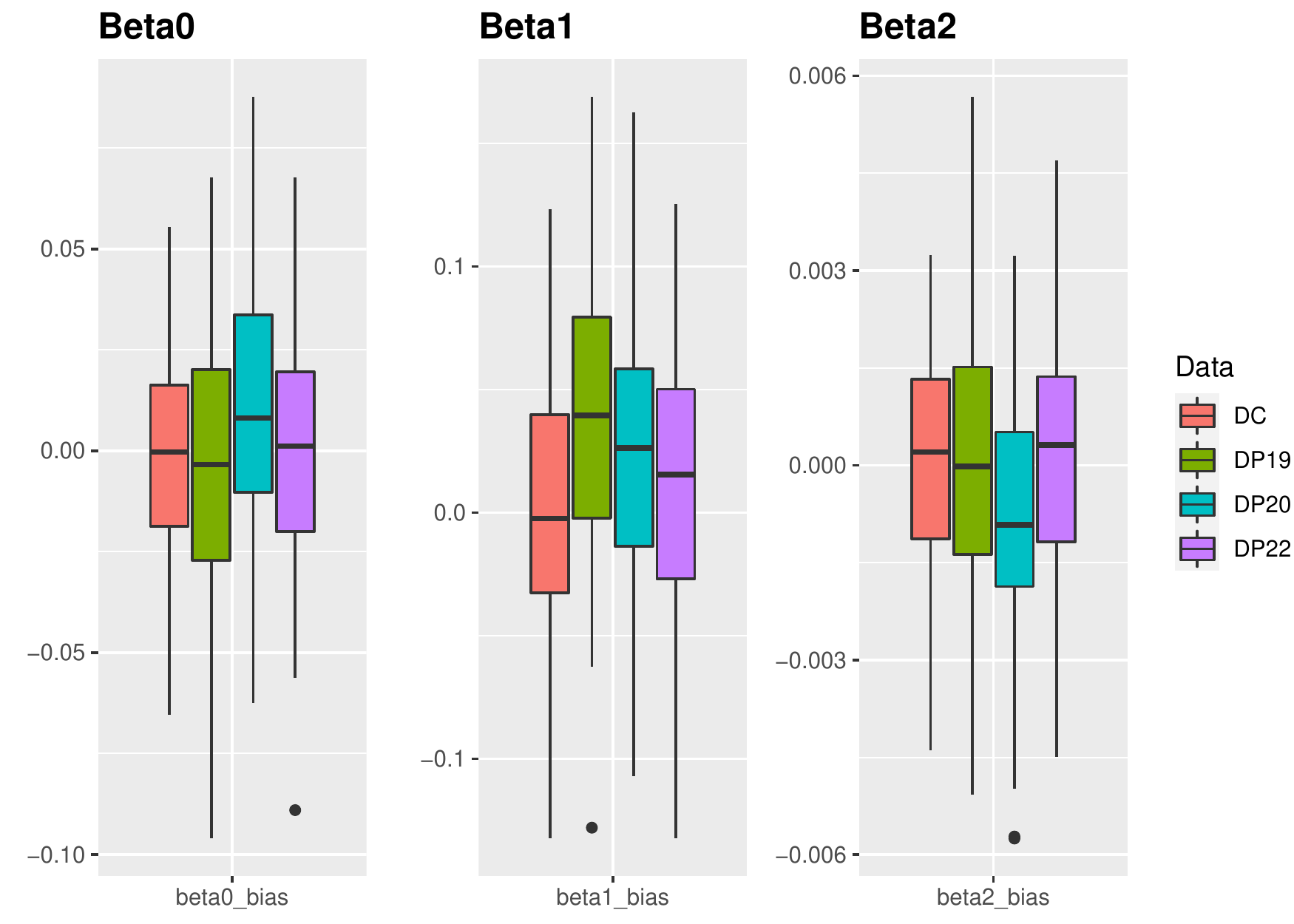} 
\label{coef_boxplot}
\end{figure}

The bias in the coefficient estimates from 100 simulated datasets are summarized in boxplots (Figure \ref{coef_boxplot}), with true parameter values $\beta_0=0, \beta_1=0.4, \beta_2=0.01$. First, we note that all three coefficient parameters are estimated with little to no bias, on average, when using the DC denominators (the "true" denominators used to generate the data) for model fitting. On the other hand, we can observe that using the DP19 and DP20 denominators in the model fitting leads to overestimation of the racialized inequity parameter ($\beta_1$), with average bias of 0.039 and 0.026, respectively, corresponding to percent biases of 10\% and 6\%, respectively.  This is likely a result of the systematic under-estimation of the DAS-protected denominators for Black populations (Figure \ref{boxplot_percent_error}), leading Black premature mortality rates to be over-estimated, a phenomenon which is not mirrored in the NHW population. This issue is largely attenuated in DP22, with an average bias of 0.009 (or 2\%) in the racialized inequity parameter estimate. The intercept and the poverty coefficient are generally estimated with little bias for all denominator data sources, although the DP20 data yield slightly more bias in these parameter estimates than the other denominators.

\begin{figure}[H] 
\caption{Boxplots of bias and mean absolute percent error (MAPE) in the standardized mortality ratio (SMR) estimates for the Black and non-Hispanic white populations in each Massachusetts census tract from models using each of the three denominator data sources.} 
\centering 
\includegraphics[width=1.0\textwidth]{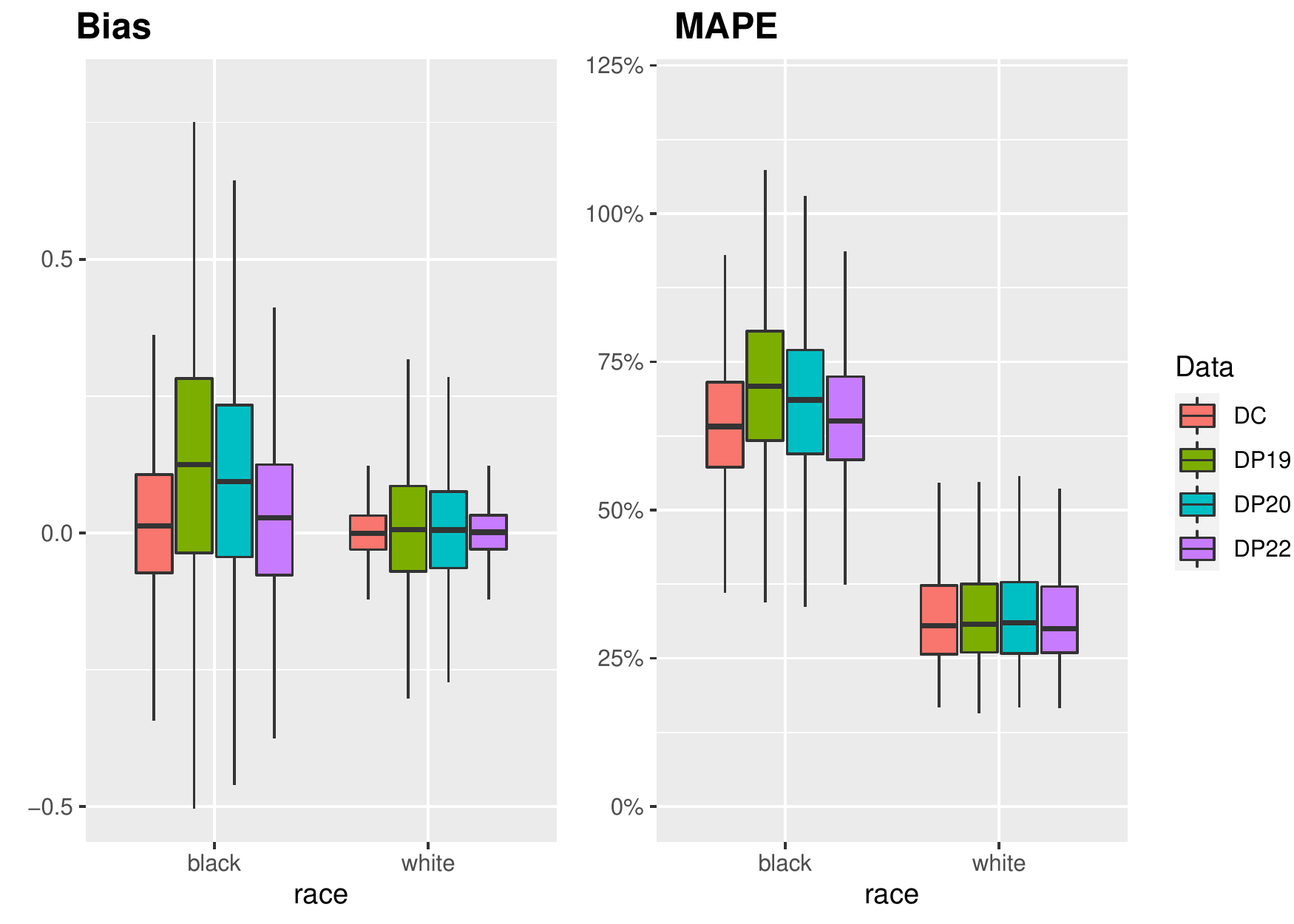} 
\label{smr_mape_bias}
\end{figure}

Figure \ref{smr_mape_bias} shows boxplots of the bias and MAPE in the model-estimated SMRs using each denominator data source. (Note that, in these plots, the data represented are the CT and racialized group-specific bias/MAPEs, averaged across simulations as in equations (3) and (4), as opposed to Figure~\ref{coef_boxplot} where the data are the difference between an estimate and the truth for each individual simulation.) When using the true DC denominators in the models, estimated SMRs are unbiased on average for both racialized groups, but the MAPEs are much larger for the Black SMRs compared to the NHW SMRs. This indicates that even correctly specified models, in absence of error in the denominators, struggle more to estimate Black vs. NHW SMRs for any given CT simply due to the smaller population sizes and more unstable rates for the Black population in most MA CTs.

The use of DP19 and DP20 denominators exacerbates this disparity in SMR estimation accuracy. With these denominators, SMR estimates for the NHW population remain unbiased on average, but for the Black population even the average of the SMR estimates demonstrates an upward bias of 0.128 and 0.090 for DP19 and DP20, respectively. This is further illustrated in Table \ref{percent_upward_bias}, which provides the percent of SMRs biased upwards for both racialized groups from each of the four denominator data sources. This is, again, the result of the systematic under-estimation of the DAS-protected denominators for Black populations (Figure \ref{boxplot_percent_error}). There is also a larger DC vs DP19/DP20 differential in the distribution of MAPEs for the Black population (relative to NHW), indicating that the use of these DAS variants worsens the (already poorer) model performance for Black populations more than for NHW populations. We also note that biases/MAPEs in the SMRs are generally smaller when using the DP20 denominators compared to the DP19 denominators. The distributions of SMR biases and MAPEs using DP22 denominators are virtually indistinguishable from those observed when using the DC denominators (average bias of 0.021 for DP22 SMRs), indicating that using the newest DAS variant for SMR estimation gives results with comparable accuracy to the gold standard.

\begin{figure}[H] 
\caption{Biases for DP20 SMRs across MA (top row) and Boston (bottom row) CTs.} 
\centering 
\includegraphics[width=1.0\textwidth]{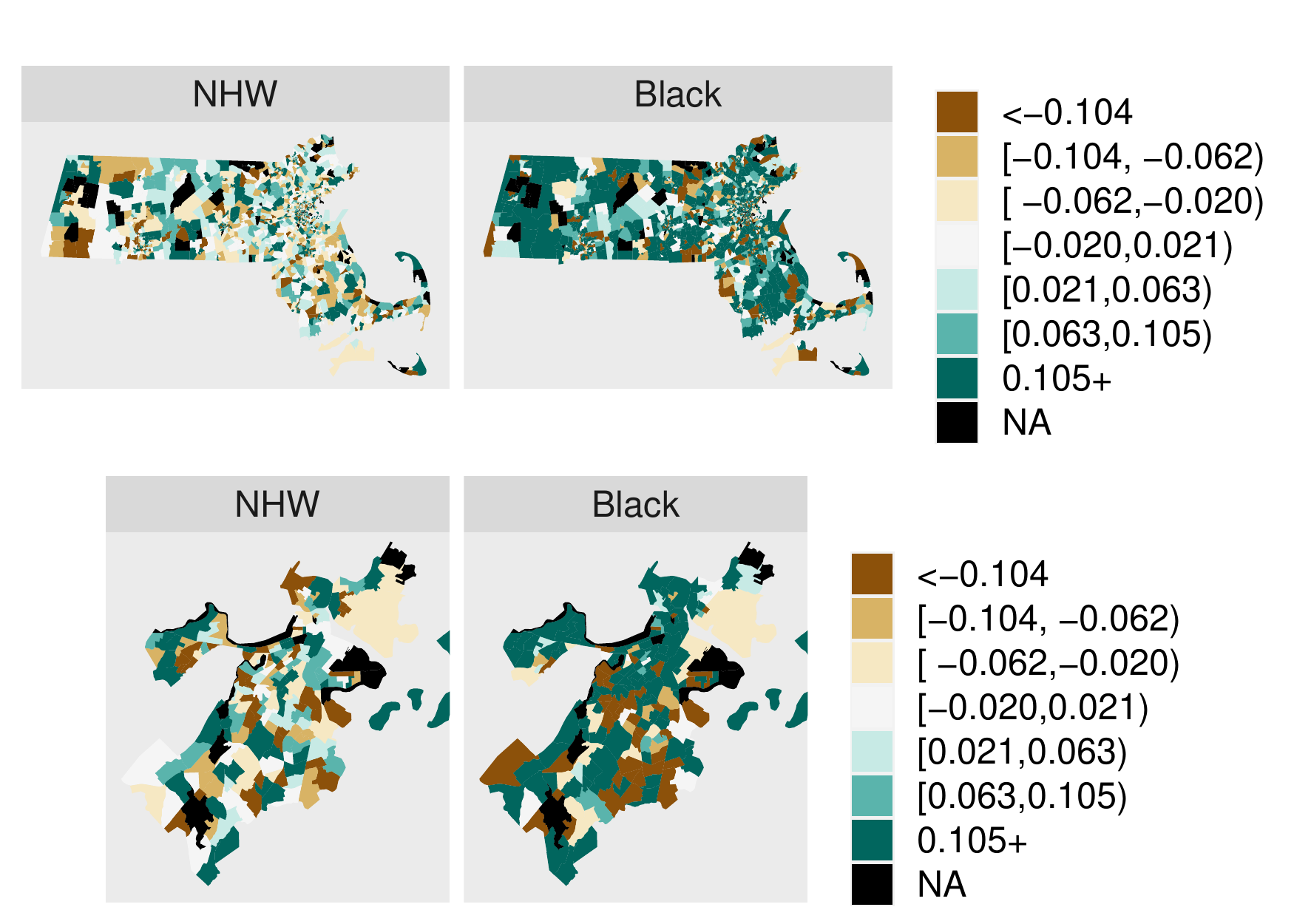} 
\label{bias_map}
\end{figure}

\begin{figure}[H] 
\caption{MAPE for DP20 SMRs across MA (top row) and Boston (bottom row) CTs.} 
\centering 
\includegraphics[width=1.0\textwidth]{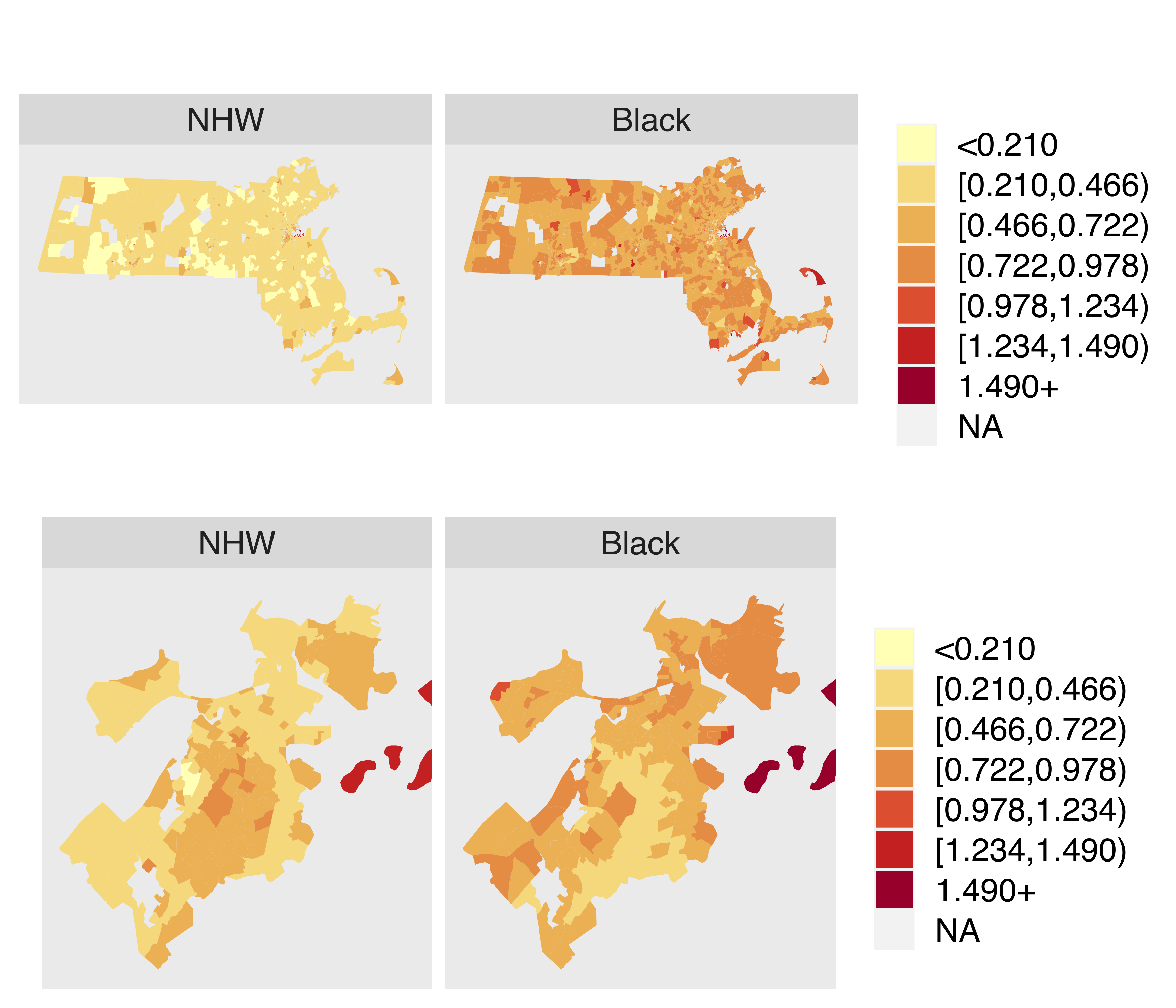} 
\label{mape_map}
\end{figure}

\begin{figure}[htb] 
\caption{Biases for DP22 SMRs across MA CTs (The first row indicates the whole area of MA, and the second row indicates the Boston area)}
\centering 
\includegraphics[width=1.0\textwidth]{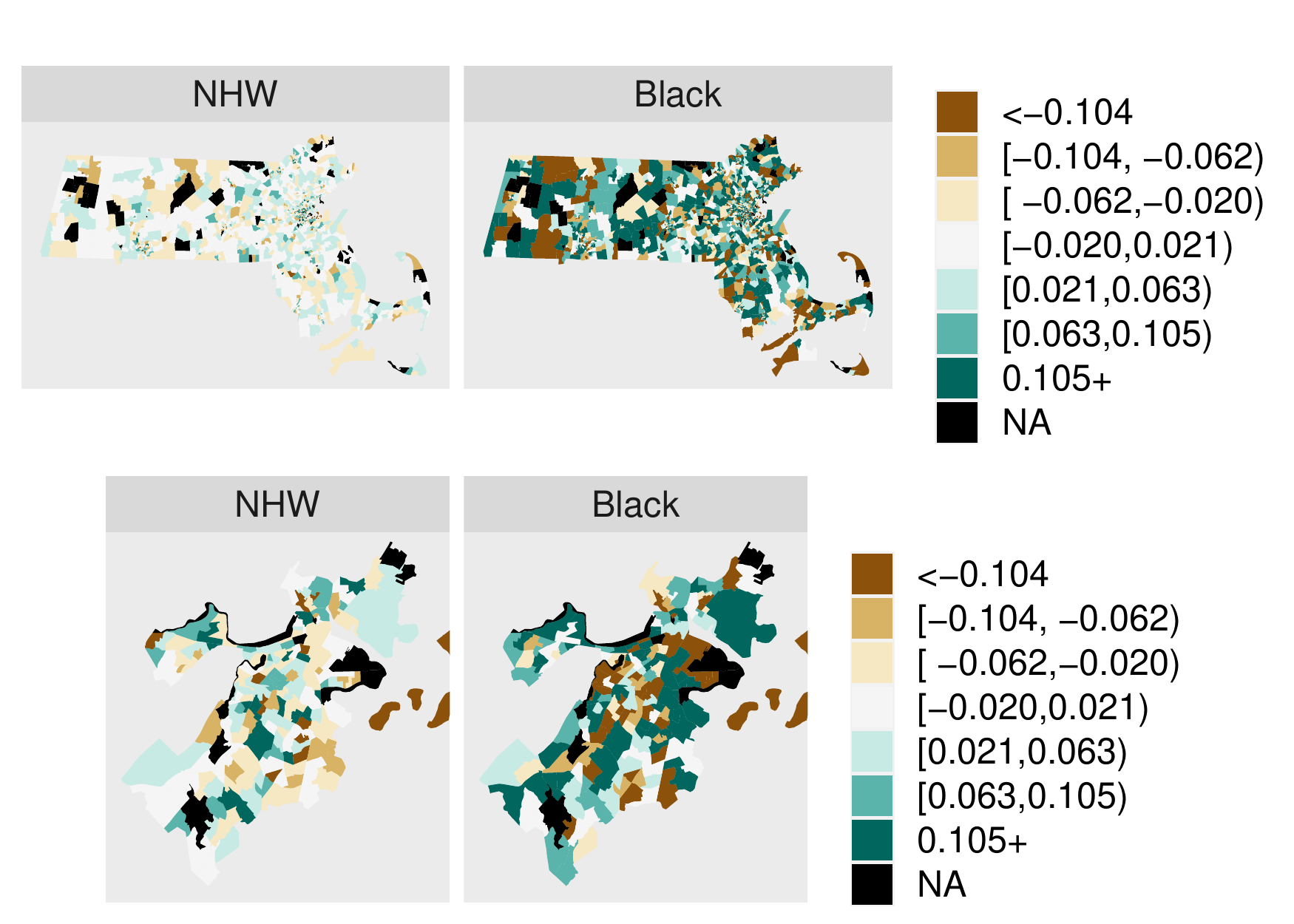} 
\label{bias_map22}
\end{figure}

\begin{figure}[htb] 
\caption{MAPE for DP22 SMRs across MA CTs  (The first row indicates the whole area of MA, and the second row indicates the Boston area)} 
\centering 
\includegraphics[width=1.0\textwidth]{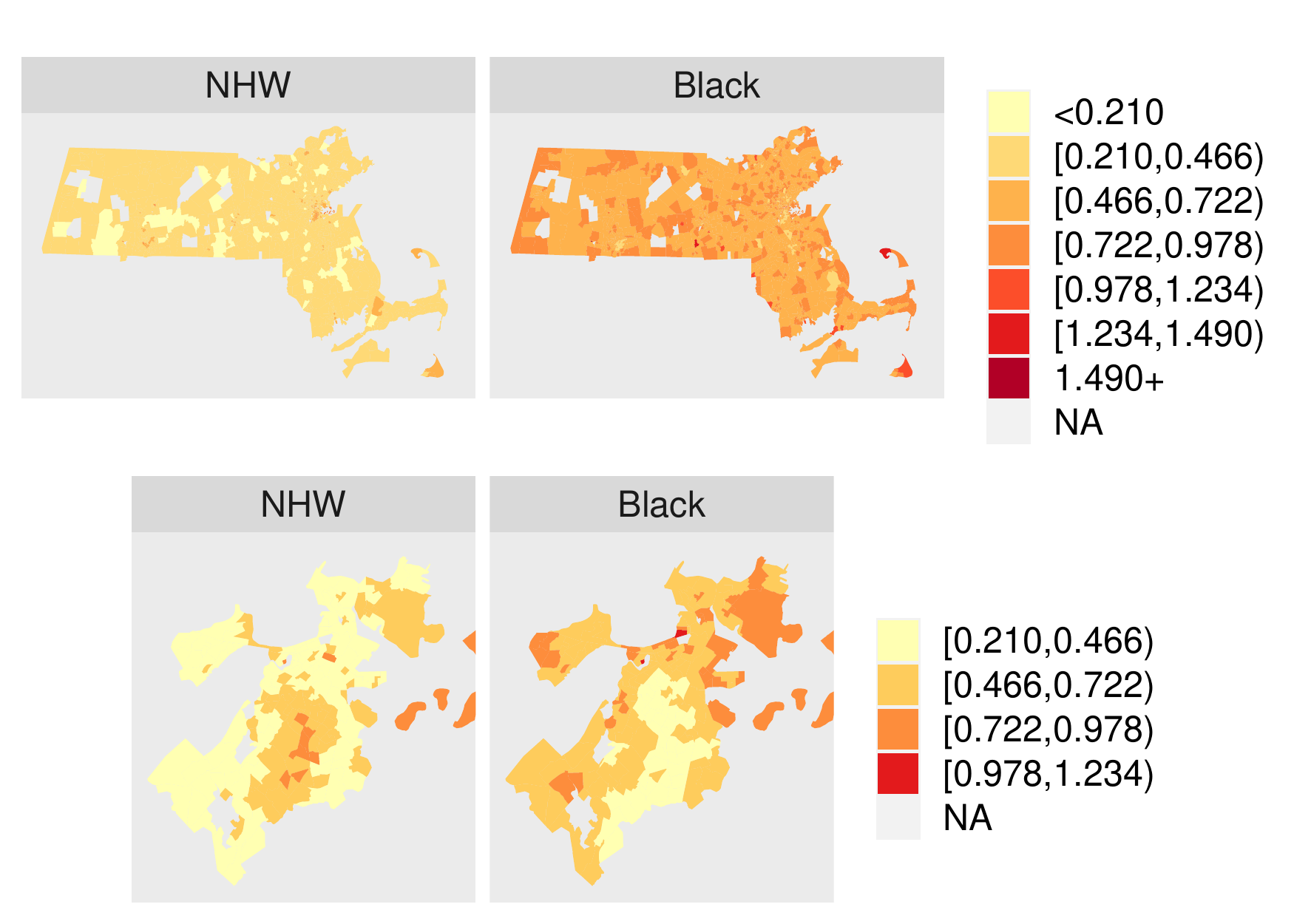} 
\label{mape_map22}
\end{figure}

The biases and MAPEs for the model-estimated SMRs for each CT and racialzed group using the DP20 denominators are mapped in Figures \ref{bias_map} and \ref{mape_map} (the patterns for DP19 are similar and the maps for DP22 are shown in Figures~\ref{bias_map22} and~\ref{mape_map22}). As noted above, the small Black populations in many MA CTs results in generally larger magnitudes of biases and MAPEs for the Black SMRs (indicated by the bolder colors in the maps). Previous studies have suggested that more severe distortions may occur when using DP denominators to characterize health-related patterns in smaller population groups\cite{doi:10.1073/pnas.200371411,doi:10.1177/2378023121994014}. Our findings, including the presence of more blue/green hues in the maps of bias in the Black SMRs (Figure \ref{bias_map}) provide further insight that the use of DP denominators using smaller PLBs tend to favor over-estimation for Black premature mortality rates but not for NHW rates (as described in Figure \ref{smr_mape_bias}). However, as we have consistently reported throughout this section, these distortions are largely eliminated by increasing the PLB to the values applied in DP22.

\section{Discussion}
In this paper, we explored the potential impact of the US Census Bureau's proposed 2020 decennial census DAS procedures, including the use of DP, taking three new steps beyond those in the handful of investigations focusing on this topic in relation to health inequities, by: (1) modeling small-area disease/mortality rates for the purposes of identifying health inequities; (2) including the newer released demonstration product that incorporates the recent Census Bureau refinements to the DAS; and (3) conducting simulation analyses to formally quantify the biases introduced by the DAS in mortality rate estimation for Black and NHW populations separately. Using three DAS-protected 2010 census demonstration products released by the Census Bureau, we conducted a simulation study and an analysis of real small-area premature mortality data from MA to investigate how the DAS procedures impacted racialized group and economic inequity estimates. 
Our results provide evidence that recent changes to the DAS procedures made by the Census Bureau in response to stakeholder feedback (featured in DP22 data)-- in particular increasing the DP PLB-- brought about substantial improvements in accuracy of the DAS-protected denominators that may translate to dramatic decreases in biases in disease mapping and inequity studies employing these denominators relative to older variants of the DAS procedures. We observed that biases in racialized and economic inequity parameter estimates and model-based SMR estimates from models using the DP22 denominators were virtually indistinguishable from those obtained when using the original DC data, indicating little impact of the DAS on analyses or inference. 

When using older demonstration products (DP19 and DP20), which applied DP with a lower PLB, we found that, while high level patterns in inequities by racialized group and socioeconomic status are preserved, the errors induced in mortality rate estimation were considerably larger for Black than NHW populations. In particular, in the example examined here, these older DAS variants led to systematic under-estimation of denominators and, therefore, over-estimation of premature mortality rates for Black populations, which was not observed for NHW populations. These relatively larger DAS-induced errors in estimated small-area SMRs for Black populations relative to NHW populations compound the already worse model performance for Black populations due to small counts and rate instability.

Our work demonstrates (1) the profound implications of the US Census Bureau's choice of PLB for the accuracy of future small-area disease mapping and health inequity studies and (2) the extent to which a small PLB, combined with modeling approaches that neglect the DAS-induced error in denominators for small groups, can distort characterizations of inequities. Our findings regarding the older DAS variants generally agree with some previous studies, which have reported that DAS-protected denominators are more problematic for estimation of rates in smaller racialized groups\cite{doi:10.1073/pnas.200371411,doi:10.1177/2378023121994014}. Such mischaracterizations could have real implications for public health practitioners, who rely on these metrics for identification of an intervention on high-risk groups. 

For instance, these distortions may lead policy makers to miss opportunities to improve public health in some small populations and/or small areas, which due to myriad social and economic factors are often the groups with the highest health risks. However, encouraging evidence from our investigation of the recent demonstration product suggests that such issues can be largely ameliorated, in this context, by the choice of a larger PLB in the DP implementation, striking a compromise between preservation of privacy and the ability to accurately characterize and advance the health of even small populations and areas.

To our knowledge, our study is so far the first study to compare different DAS-protected 2010 census demonstration products' impacts for small-area disease modeling and inequity studies and the first health-focused study to investigate the demonstration product newly released in 2022. The primary weakness of our study is that we only investigate DAS impacts in a single state, MA, and on health inequity estimates for the two largest racialized groups in MA, Black and NHW. The DAS-attributable errors uncovered here may be further exacerbated for other smaller groups, such as Native Americans and Asian Americans/Pacific Islanders. Moreover, our simulation study is conducted utilizing the DC denominators as the true denominators in the data generating process. However, even the original DC population counts are known to exhibit systematic biases, with a particular tendency to under-represent non-white individuals, which is particularly troublesome for health inequity studies\cite{o2019differential}. In spite of these limitations, DC denominators are commonly used as a ``gold standard'' for comparison purposes when evaluating alternative denominator data sources\cite{Krieger2021,NETHERY2021100786}.

As the implementation of DP to preserve privacy in publicly released health and social science data accelerates, the development of statistical methods to adapt standard disease mapping models to DP-injected noise in variables is critical to reduce DP-related systematic errors that can bias health inequity estimates. Future work should also investigate how the Census Bureau's 2020 DAS impacts health inequity studies using smaller populations, such as Native Americans and Asian Americans/Pacific Islanders.

\bibliographystyle{unsrt}  
\bibliography{references}  

\section*{Acknowledgements}
\subsection*{Funding:}
The authors gratefully acknowledge funding from NIH grants R01HD092580, 1K01ES032458 and P30ES000002.

\subsection*{Author contributions:}

YL: Conceptualization; Data curation; Formal analysis; Methodology; Visualization; Writing - original draft; Writing - review \& editing 

RCN: Conceptualization; Data curation; Formal analysis; Methodology; Visualization; Writing - original draft; Writing - review \& editing 

BAC: Conceptualization; Funding acquisition; Methodology; Writing - review \& editing 

NK: Conceptualization; Funding acquisition; Writing - review \& editing 

EP: Conceptualization; Methodology; Resources; Writing - review \& editing 

LW: Conceptualization; Funding acquisition; Methodology; Writing - review \& editing 

JTC: Conceptualization; Methodology; Resources; Writing - review \& editing

\subsection*{Competing interests:}
Authors declare that they have no competing interests.

\subsection*{Data and materials availability:}
The real premature mortality data used herein can be obtained upon request to the MA Department of Public Health. Demonstration data products are publicly available on the IPUMS website (\url{https://www.nhgis.org/privacy-protected-2010-census-demonstration-data}). 2010 decennial census data was imported using the \texttt{tidycensus} R package. Code to reproduce analyses and simulation studies is available on Github at \url{https://github.com/Lyric98/CT_race_diff_privacy}.

\section*{Figures and Tables}

\newpage

\begin{table}[H]
\caption{Mortality rate ratio estimates (95\% credible intervals) from real premature mortality data analyses with each of the four denominator data sources.}
  \label{IRR-table}
  \centering
\begin{tabular}{rllll}
 \toprule
   \multicolumn{5}{c}{Data Sources}                \\   
    \cmidrule(r){2-5}
 & DC & DP19 & DP20 & DP22 \\ 
  \midrule
Intercept & 1.06 (1.04,1.09) & 1.06 (1.04,1.09) & 1.06 (1.04,1.09) & 1.06 (1.04,1.09) \\ 
  Racialized group & 1.16 (1.06,1.29) & 1.17 (1.06,1.29) & 1.16 (1.05,1.28) & 1.16 (1.07,1.3) \\ 
  Poverty & 1.33 (1.28,1.38) & 1.33 (1.28,1.38) & 1.32 (1.28,1.37) & 1.34 (1.29,1.38) \\ 
   \bottomrule
\end{tabular}
\end{table}


\newpage

\appendix
\counterwithin{figure}{section}
\counterwithin{table}{section}

\section{Supplementary Materials}
\label{Supplementary}

\begin{table}[ht]
\caption{Percent of expected counts that are 0 (by racialized group) and the percent that are less than 5 (by racialized group) for each denominator data source}
  \label{expected_counts_percent-table}
  \centering

  
\begin{tabular}{rllll|llll}
 \toprule
   && NHW &&& Black\\
    \cmidrule(r){2-9}
Data Sources & DC & DP19 & DP20 & DP22 & DC & DP19 & DP20 & DP22\\ 
  \midrule
Percent of 0 expected counts & 0.00 & 0.00 & 0.07 & 0.00 & 0.14 & 1.64 & 1.37 & 0.27\\ 
  Percent of $<$5 expected counts & 39.32 & 38.90 & 39.32 & 39.25 & 100.00 & 100.00 & 100.00 & 100.00\\ 
   \bottomrule
\end{tabular}
\end{table}

\begin{table}[ht]
\caption{Summary statistics (Min, Max, 25th, 50th, and 75th percentiles) of the difference between both DP datasets and DC dataset for Black and NHW separately}
  \label{summary_dp_minus_dc}
  \centering
\begin{tabular}{rrlllll}

 \toprule

 && 0\% & 25\% & 50\% & 75\% & 100\% \\ 
  \midrule
\multirow{2}{*}{DP19}
& NHW  & -2.6348 & -0.3293 & -0.0026 & 0.3548 & 1.7335 \\
& Black & -0.9207 & -0.0515 & -0.0163 & 0.0371 & 1.1627 \\
  \midrule

\multirow{2}{*}{DP20}
& NHW& -1.8218 & -0.3212 & -0.0049 & 0.3066 & 1.6439 \\
& Black & -0.4807 & -0.0404 & -0.0093 & 0.0245 & 0.6865\\  
\midrule

\multirow{2}{*}{DP22}
& NHW& -0.1258 & -0.0210 & 0.0001 & 0.0228 & 0.1097 \\
& Black & -0.1090 & -0.0125 & -0.0009 & 0.0123 & 0.0764\\   \bottomrule
\end{tabular}
\end{table}

\begin{table}[ht]
\caption{Percent of under-estimated expected DP counts}
  \label{expected_counts_percent-table}
  \centering
\begin{tabular}{rrrr}
  \hline
 &  DP19  &  DP20  & DP22 \\ 
  \hline
NHW & 50.3084 & 50.3770 & 49.6916 \\ 
  Black & 62.7645 & 59.3794 & 51.9746 \\ 
   \hline
\end{tabular}
\end{table}

\begin{table}[ht]
\caption{Average biases in the coefficient estimates from 100 simulated datasets from all four data sources.}
  \label{IRR-table}
  \centering
\begin{tabular}{rllll}
 \toprule
   \multicolumn{5}{c}{Data Sources}                \\   
    \cmidrule(r){2-5}
 & DC & DP19 & DP20 & DP22 \\ 
  \midrule
Intercept & -0.0006 & -0.0046 & 0.0114 & -0.0005 \\ 
Racialized group CT & 0.0016 & 0.0386 & 0.0258 & 0.0093 \\ 
  Proverty & 0.0001 & 0.0001 & -0.0008 & 0.0001 \\ 
   \bottomrule
\end{tabular}
\end{table}

\begin{table}[ht]
\caption{Percent of upward bias for SMRs}
  \label{percent_upward_bias}
  \centering
\begin{tabular}{rrrrr}
  \hline
 & DC &  DP19  &  DP20  & DP22 \\ 
  \hline
NHW & 49.3489 & 52.0905 & 51.8849 & 51.5422 \\ 
  Black & 54.0197 & 70.2398 & 68.3357 & 57.7574\\ 
   \hline
\end{tabular}
\end{table}

\begin{table}[ht]
\caption{Average biases for SMRs}
  \label{avg_bias}
  \centering
\begin{tabular}{rrrrr}
  \hline
 & DC &  DP19  &  DP20  & DP22 \\ 
  \hline
NHW & 0.0006 & 0.0164 & 0.0104 & 0.0012 \\ 
  Black & 0.0049 & 0.1281 & 0.0899 & 0.0209\\ 
   \hline
\end{tabular}
\end{table}

\begin{table}[ht]
\caption{Average MAPEs for SMRs}
  \label{avg_mape}
  \centering
\begin{tabular}{rrrrr}
  \hline
 & DC &  DP19  &  DP20  & DP22 \\ 
  \hline
NHW & 0.3313 & 0.3412 & 0.3391 & 0.3323 \\ 
  Black & 0.6447 & 0.7123 & 0.6886 & 0.6548\\ 
   \hline
\end{tabular}
\end{table}

\begin{figure}[htb] 
\caption{Scatterplots of 2010 decennial census (DC) expected premature mortality counts (x-axis) versus 2010 demonstration product expected counts (y-axis) for MA census tracts for the Black and non-Hispanic white populations.} 
\centering 
\includegraphics[width=0.8\textwidth]{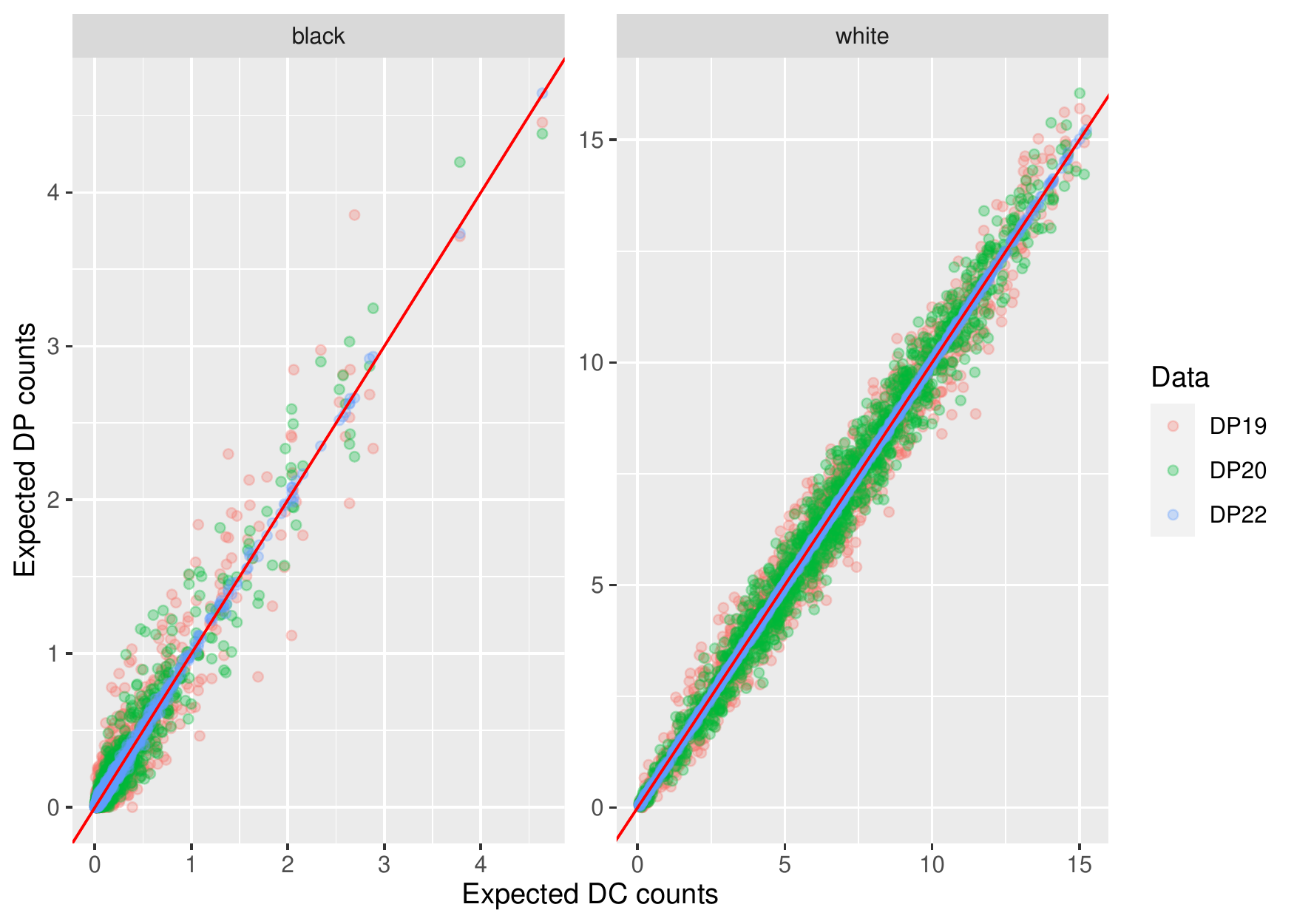} 
\label{scatterplot}
\end{figure}

\end{document}